\documentclass{jpsj3}
\usepackage{txfonts}
\usepackage{braket}
\usepackage{color}
\usepackage{dcolumn}
\usepackage{bm}

\title{Incommensurate Magnetic Ordered Phase with Enhanced Low-Temperature Magnetic Specific Heat in SmAu$_3$Al$_7$}

\author{Ryuji Higashinaka\thanks{E-mail address: higashin@tmu.ac.jp}$^{,\ 1}$, Takuma Iwami$^{1}$, Kohsuke Saitou$^{1}$, Takashi U. Ito$^{2}$, Chihiro Tabata$^{2, 3}$, Koji Kaneko$^{2, 3}$, Takashi Ohhara$^{4}$, Ryoji Kiyanagi$^{4}$, Akiko Nakao$^{5}$, Jumpei G. Nakamura$^6$, Wataru Higemoto$^{2, 4, 7}$, Akihiro Koda$^6$,  Shinsaku Kambe$^2$, Yuji Aoki$^{1}$, and Tatsuma D. Matsuda$^{1}$}
\inst{$^{1}$Department of Physics, Tokyo Metropolitan University, Hachioji-shi, Tokyo 192-0397, Japan\\
$^{2}$Advanced Science Research Center, Japan Atomic Energy Agency, Tokai, Ibaraki 319-1195, Japan\\
$^{3}$Materials Sciences Research Center, Japan Atomic Energy Agency, Tokai, Ibaraki 319-1195, Japan\\
$^{4}$J-PARC Center, Japan Atomic Energy Agency, Tokai, Ibaraki 319-1195, Japan\\
$^{5}$Comprehensive Research Organization for Science and Society, Tokai, Ibaraki 319-1106, Japan\\
$^{6}$Institute of Materials Structure Science, High-Energy Accelerator Research Organization, Tsukuba, Ibaraki 305-0801, Japan\\
$^{7}$Department of Physics, Institute of Science Tokyo, O-okayama, Meguro, Tokyo 152-8551, Japan
}
\abst{
Neutron scattering and muon spin rotation ($\mu$SR) measurements on single-crystal SmAu$_3$Al$_7$ reveal magnetically ordered states associated with successive transitions at $T_{\rm N}$ = 2.8 K and $T^*$ = 0.9 K. 
Magnetic Bragg peaks appear below $T_{\rm N}$ with an incommensurate (IC) propagation vector ${\bf q}$ = (0.30, 0, 1.33). 
$\mu$SR detects spontaneous internal fields below $T_{\rm N}$, and the spectral shape is consistent with the IC magnetic ordering. 
No anomalies are observed at $T^*$, indicating that the magnetic structure remains essentially unchanged below and above $T^*$. 
The magnetic order is revealed to be a spatially homogeneous long-range ordered state, rather than a partially disordered state proposed in earlier studies.
The possible connection between the IC magnetic order and the enhanced low-temperature magnetic specific heat is discussed.
}
\date{\today}
\begin{document}
\maketitle
%
 Recent studies have revealed a variety of unique strongly correlated electronic properties inherent to Sm systems. 
These include an unusual magnetic-field-insensitive heavy-fermion behavior observed in SmOs$_4$Sb$_{12}$ \cite{Sanada_JPSJ_05}, magnetic-field-insensitive phase transitions accompanied by significantly enhanced Sommerfeld coefficients ($\gamma$) in Sm{\it Tr}$_2$Al$_{20}$ ({\it Tr} = Ti, V, Cr, and Ta) \cite{Higashinaka_JPSJ_11,Yamada_JPSJ_13,Sakai_PRB_11}, and a partially disordered Sm-ion state in SmPt$_2$Si$_2$ \cite{Fushiya_JPSJ_14}.
Recently, we have identified SmAu$_3$Al$_7$ as a candidate for realizing a partial Kondo screening (PKS) state \cite{Higashinaka_JPSJ_23}. 
SmAu$_3$Al$_7$ crystallizes in the $R{\bar 3}c$ ScRh$_3$Si$_7$-type structure, where Sm ions are encapsulated within cages possessing ${\bar 3}$ symmetry along the $c$-axis of the hexagonal representation \cite{Latturner_JSSC_03,Higashinaka_JPSCP_20}, as illustrated in Fig. \ref{Crystal}(a).
This compound undergoes successive phase transitions at $T_{\rm N}$ = 2.8 K and $T^*$ = 0.9 K, giving rise to two ordered phases: phase I ($T^* < T < T_{\rm N}$) and phase II ($T < T^*$), as shown in Fig. \ref{Crystal}(b). 
The $c$-axis magnetic susceptibility, characterized by pronounced Ising anisotropy, exhibits a minor decrease just below $T_{\rm N}$, indicative of antiferromagnetic (AFM) ordering. 
This decrease, however, is subtle, and the susceptibility increases again within the magnetically ordered phase I [Fig. \ref{Crystal}(c)].
An analysis of the effective magnetic moment in phase I suggests that approximately 40 \% of the Sm magnetic moments remain in a paramagnetic state in this temperature range. 
Furthermore, a pronounced enhancement of the electronic specific heat coefficient, exceeding 1 J/(mol K$^2$), is observed in phase II. 
These findings indicate that the residual degrees of freedom associated with the Sm 4{\it f} electrons may play a crucial role in forming the heavy-fermion state \cite{Higashinaka_JPSJ_23}; however, this possibility has yet to be conclusively verified.

In this study, we performed neutron scattering and muon spin rotation ($\mu$SR) measurements using single crystals to elucidate the peculiar magnetically ordered ground state of this compound and to gain further insight into the possible PKS state. 
Despite the strong neutron absorption of Sm, several magnetic reflections were successfully observed. 
The result revealed an incommensurate (IC) magnetic order with the magnetic propagation vector ${\bf q}$ = (0.30, 0, 1.33), while no clear change was observed on passing through $T^*$. 
In the $\mu$SR measurements, we observed the existence of the spontaneous local field below $T_{\rm N}$, indicating the magnetic ordering. 
These results can be consistently explained by IC magnetic ordering, in agreement with the neutron scattering findings.

Single crystals of SmAu$_3$Al$_7$ were grown by the Al self-flux method \cite{Higashinaka_JPSJ_23}.
The isotopic composition of Sm in the present samples corresponds to the natural abundance.

Neutron scattering measurements were carried out using the single-crystal time-of-flight (TOF) neutron diffractometer SENJU\cite{SENJU}, installed at BL-18 of the Materials and Life Science Experimental Facility (MLF), J-PARC, and the thermal-neutron triple-axis spectrometer TAS-1, installed at the 2G beam port of JRR-3, JAEA, Tokai, Japan. 
In SENJU, two setups, $\lambda$ with 0.4-4.4 \AA and 4.6-8.8 \AA, were employed to cover wide reciprocal lattice space. 
The sample was cut to have a large $c$-plane surface with dimensions of 5 $\times$ 5 $\times$ 0.5 mm$^3$ in order to have wide coverage in the ($h$ 0 $l$) scattering plane. 
The crystal was glued to a copper plate and mounted of the cold finger of a $^3$He cryostat.
Diffraction intensities were collected at two crystal orientations and four temperatures: 0.3 K (phase II), 1.2 K (phase I), and 3.5 K and 6 K (paramagnetic (PM) phase), with an exposure time of approximately 6 h for each configuration.
Data reduction and visualization were performed using the STARGazer software package. \cite{STARGazer}
The UB matrix was refined by using the data at 0.3 K. 

After finding magnetic peaks at SENJU, further investigation was performed on the TAS-1 spectrometer with the same sample in the ($h$ 0 $l$) scattering plane.
A neutron wavelength of 1.41 \AA\ was employed to reduce neutron absorption. 
The temperature dependence of the magnetic scattering intensity was measured between 0.3 K and 3.5 K using a $^3$He refrigerator.

$\mu$SR measurements on single-crystalline SmAu$_3$Al$_7$ were performed at the S1 area of J-PARC MUSE.
Fourteen single-crystal pieces were mounted on a silver sample holder with their $c$ axes aligned parallel to the muon-beam direction, while their in-plane orientations were random.
The $\mu$SR experiments were conducted under zero field (ZF) and under a weak transverse field (wTF) of 2 mT applied perpendicular to the $c$ axis, in the temperature range of 0.3-5.1 K using a $^3$He refrigerator.
All data were analyzed using the musrfit software suite \cite{musrfit}.

Figure \ref{SENJU}(a) shows the neutron scattering intensity map of the ($h$ 0 $l$) plane for SmAu$_3$Al$_7$ at 0.3 K, corresponding to phase II, measured on SENJU.
Despite the strong neutron absorption of Sm, nuclear Bragg peaks, indicated by arrows in the figure, were clearly observed.
The sharpness of these Bragg peaks demonstrates the high crystalline quality of the present single-crystal sample.
In addition to the nuclear Bragg peaks, weak but distinct superlattice reflections—marked by circles in Fig. \ref{SENJU}(a)—were detected at four equivalent reciprocal lattice positions, each described by the propagation vector ${\bf q} \sim (0.3, 0, 1.33)$. 
No magnetic scatterings were observed outside the ($h$ 0 $l$) plane, and no higher-harmonic components of the propagation vector q were detected within the experimental accuracy.
Figures \ref{SENJU}(b) and \ref{SENJU}(c) display neutron scattering intensity maps obtained at 1.2 K (phase I) and 6 K (PM phase), respectively, within the reciprocal-space region enclosed by the dotted square in Fig. \ref{SENJU}(a). 
The superlattice reflections persist at the same reciprocal positions in phase I with reduced intensity, and disappear in the PM phase. 
The appearance of these reflections only below $T_{\rm N}$ clearly indicates that they correspond to magnetic Bragg peaks arising from long-range magnetic order.
Furthermore, the positions of the magnetic reflections remain unchanged between phases I and II, implying that the propagation vector of the magnetic order does not vary across $T^*$.

To precisely determine the magnetic propagation vector ${\bf q}$ and to evaluate the order parameter below $T_{\rm N}$, single-crystal neutron diffraction experiments were performed on the TAS-1 spectrometer.
Because magnetic Bragg reflections generally exhibit stronger intensities at smaller ${\bf Q}$ values, a search for superlattice reflections was conducted in the low-${\bf Q}$ region.
A distinct superlattice reflection was identified at the reciprocal-lattice position (1.7, 0, 0.66).
Scans along the $h$ and $l$ directions in reciprocal space were carried out around this position at 0.3 K (phase II), as shown in Fig.~\ref{QVec}. 
Gaussian fitting of the measured intensity profiles at several superlattice positions revealed that the magnetic propagation vector is ${\bf q} = (0.30, 0, 1.33)$, indicating an IC magnetic order that remains unchanged between phases I and II (data not shown).
The superlattice reflections were observed only at reciprocal-lattice positions $\pm(0.30, 0, 1.33)$ relative to the nuclear Bragg peaks.

In addition, to obtain information on the magnetic structures in phases I and II, the intensity ratios of the several magnetic reflections were compared. 
Figure 4 shows the $\theta$-2$\theta$ scan profiles at (2, 0, 2)-${\bf q}$, (2, 0, 2)+${\bf q}$, and (2, 0, -4)+${\bf q}$ measured at 0.3 K and 1.2 K. 
For each reflection, the ratio of the integrated intensities at 0.3 K and 1.2 K, $I_{\rm 1.2\, K}/I_{\rm 0.3\, K}$, was approximately 0.6, showing no ${\bf Q}$ dependence. 
These results indicate that, within the limits of experimental accuracy, the magnetic structure—specifically its orientation and modulation—remains unchanged between phases I and II, and only the magnitude of the ordered moment varies with temperature.

The temperature dependence of the scattering intensity at the superlattice reflection with the highest peak intensity was measured.
The intensity begins to increase below $T_{\rm N}$, while no distinct anomaly is observed at $T^*$.
This result will be discussed later in relation to the $\mu$SR measurements.

The transition at $T^*$ is of first-order character, as indicated by the hysteresis observed in the temperature dependence of the specific heat upon heating and cooling \cite{Higashinaka_JPSJ_23}. 
To examine whether the transition at $T^*$ is associated with a structural change, $\theta$-2$\theta$ scans of the nuclear Bragg reflections were performed in the PM, phase I, and phase II states (data not shown).
The absence of any change in the peak positions or profiles of the nuclear Bragg reflections across $T_{\rm N}$ and $T^*$ confirms that there is no significant variation in the lattice constants or lowering of crystal symmetry at either transition temperature within the present sensitivity.

To gain further insights into the magnetic properties of the ordered phase of this compound, $\mu$SR measurements were performed. 
Figure~\ref{ZF_OP}(a) shows the ZF-$\mu$SR spectra of SmAu$_3$Al$_7$ measured at 5.1, 2.0, and 0.31 K. 
The background contribution from the silver cold plate was subtracted.
In the PM phase above $T_{\rm N}$ = 2.8 K, the ZF-$\mu$SR signal exhibits a depolarization toward zero. 
Below $T_{\rm N}$, muon spin precession is observed, indicating the development of a spontaneous local magnetic field at the muon site in phase I.
This result demonstrates that the primary order parameter in phase I is of magnetic-dipole origin. 
To estimate the magnitude and temperature evolution of the spontaneous local magnetic field associated with the precession, the time spectra were analyzed using the following functions:
\begin{equation}
A(t) = A_1 \exp(-\lambda_1 t) + A_2 J_0 \left( 2\pi f t \right) \exp(-\lambda_2 t), 
\end{equation}
where $A_1$ ($\lambda_1$) and $A_2$ ($\lambda_2$) denote the partial asymmetry (depolarization rate) corresponding to the longitudinal and transverse components, respectively.
Because IC magnetic ordering is expected from the neutron results, a zeroth-order Bessel function $J_0$ was employed to describe the oscillating component\cite{Bessel_analysis_muSR}.
The frequency $f$ is related to $f = \gamma_\mu B_{\rm max}/2\pi$, where $\gamma_\mu = 2\pi \times 135.53$ MHz/T is the muon gyromagnetic ratio and $B_{\rm max}$ corresponds to the high-field cutoff of the continuous distribution of local magnetic fields arising from the IC magnetic order. 

The resulting temperature dependence of $B_{\rm max}$ is shown in Fig.~\ref{ZF_OP}(b).
With decreasing temperature, $B_{\rm max}$ starts to increase below approximately 2.8 K and saturates at around 9 mT.
A scaling relationship was found between $B_{\rm max}$ obtained from the $\mu$SR and the square root of the magnetic scattering intensity from neutron diffraction $I_{\rm mag}$, which is proportional to the magnitude of ordered moment. 
The normalized values of $B_{\rm max}$ and $\sqrt{I_{\rm mag}}$ are plotted as functions of temperature in Fig.~\ref{ZF_OP}(b), showing a good agreement between them. 
The order-parameter behavior is well fitted by a gap function, as expected for a transition associated with gap opening in $k$-space, such as a spin-density-wave transition \cite{Overhauser_PR_62, Fawcett_RMP_88}. 
This consistency indicates that both neutron and $\mu$SR measurements probe the same magnetic ordering of Sm.
In contrast to the clear increase at $T_{\rm N}$, no discernible anomaly is observed at $T^*$.
This suggests that any change in the ordered magnetic moment and/or magnetic structure at $T^*$ is too small to be resolved in the present experiments.

To evaluate the magnetically ordered volume fraction following the method described in Ref.~\cite{wTF_analysis}, wTF-$\mu$SR measurements were performed under an applied field of 2 mT.
The details of the analysis are found in supplemental material.\cite{SM1} 
The results indicate that the magnetic order develops homogeneously throughout the entire sample volume, ruling out any macroscopic phase separation between magnetically ordered and PM regions.

We now discuss the origin of the reemergence of the Curie term below $T_{\rm N}$ observed in the temperature dependence of the magnetic susceptibility.
The present study has revealed that an IC long-range magnetic order sets in at $T_{\rm N}$.
In this case, there are possibly no distinct Sm sites that separately exhibit magnetic order and PM state; rather, all Sm sites participate uniformly in the IC magnetic order. 
An increase in magnetic susceptibility within an IC magnetically ordered phase, as observed in the present compound, has also been reported for UPd$_2$Si$_2$ and successfully interpreted in terms of the axial next-nearest-neighbor Ising (ANNNI) model within the mean-field approximation \cite{Honma_JPSJ_98}.
A similar mechanism may be relevant to the present system.

Next, we discuss the enhancement of the specific heat at low temperatures.
In many magnetically ordered systems, an IC magnetic structure is stable in the temperature region just below the magnetic transition, while at lower temperatures the system typically undergoes a transition to a commensurate anti-phase (square) structure to release magnetic entropy.
In contrast, SmAu$_3$Al$_7$ exhibits a phase transition at $T^*$, but the magnetic structure remains essentially unchanged across this transition, and the propagation vector shows no detectable variation down to 0.3 K.
Similar behavior has been reported in several rare-earth strongly correlated systems, where it has been proposed that a fraction of the local 4{\it f}-electron degrees of freedom remains active at very low temperatures due to Kondo screening arising from 4{\it f}-conduction electron hybridization \cite{Vettier_PRL_86, Bonville_EPL_00, Onimaru_PRL_05}.
By analogy, it is plausible that such residual 4{\it f}-electron degrees of freedom contribute to the phase transition at $T^*$ and to the enhancement of the electronic specific heat coefficient below $T^*$.
However, the present experiments have not yet clarified the detailed mechanism.
As observed in SmAu$_3$Al$_7$, a pronounced enhancement of the electronic specific heat coefficient in the magnetically ordered phase has also been reported for several Sm-based strongly correlated systems, including the Sm$Tr_2$Al$_{20}$ family ({\it Tr} = Ti, V, Cr, and Ta) \cite{Higashinaka_JPSJ_11, Yamada_JPSJ_13, Sakai_PRB_11} and SmPt$_2$Si$_2$ \cite{Fushiya_JPSJ_14}.
This behavior may represent a characteristic feature of Sm-based correlated electron systems, reflecting the dual localized-itinerant nature of the Sm 4{\it f} electrons.
Very recently, de Haas-van Alphen measurements on SmTi$_2$Al$_{20}$ have revealed a Fermi surface with an extremely large cyclotron mass—approximately 26 times the bare electron mass—within the magnetically ordered phase \cite{Asif_JPSJ_24}.
These findings further support the possibility of heavy-fermion formation associated with the residual 4{\it f}-electron degrees of freedom.
Clarifying the microscopic origin of this unconventional enhancement of the electronic specific heat in the magnetically ordered state remains an important subject for future investigation.

In summary, the magnetic Bragg reflections arising from the ordering of Sm magnetic moments were successfully observed in a bulk single crystal without Sm isotope enrichment.
By combining neutron scattering and $\mu$SR measurements, we have demonstrated that SmAu$_3$Al$7$ exhibits a spatially homogeneous IC magnetic order with a propagation vector ${\bf q} = (0.30, 0, 1.33)$ below $T_{\rm N}$, rather than a partially disordered phase.
The propagation vector shows no noticeable temperature dependence and remains unchanged below $T^*$.
The observed {\bf Q} dependence of the magnetic scattering intensity indicates no detectable change in the magnetic structure across $T^*$. 
Furthermore, the temperature dependences of the order parameters obtained from both techniques follow the same characteristic curve.

\begin{acknowledgments}
The neutron diffraction experiments on SENJU at MLF J-PARC and on TAS-1 at JRR-3 JAEA were performed under a user program (Proposal No. 2022A0194, and D665), respectively.
The muon spin relaxation measurements were performed on S1 at MLF J-PARC under a user program (Proposal No. 2020B0357). 
This work was partly supported by the Japanese Ministry of Education, Culture, Sports, Science, and Technology (MEXT), Japan Society for the Promotion of Science (JSPS) Kakenhi Grant-in-Aid (Numbers: JP20H01864, JP22K03517, JP23H04867, JP23H04870, JP25K07228) and the Tokyo Metropolitan Government Advanced Research Grant Number (H31-1).
\end{acknowledgments}

\clearpage
\begin{figure}[ht]
	\begin{center}
		\caption{
		(Color online) (a) Crystal structure of SmAu$_3$Al$_7$ in the hexagonal representation, drawn using the VESTA program \cite{VESTA}.
		(b) and (c) Temperature dependence of the magnetic specific heat, $C_{\rm mag}/T$, and magnetic susceptibilities, reproduced from Ref.~\cite{Higashinaka_JPSJ_23}.
		}
	\label{Crystal}
	\end{center}
\end{figure}
\begin{figure}[ht]
	\begin{center}
		\caption{
		(Color online) (a) Neutron scattering intensity map of the reciprocal ($h$ 0 $l$) plane of SmAu$_3$Al$_7$ at 0.3 K (phase II), extracted from diffraction data collected on SENJU.
		Reflections indicated by arrows and circles correspond to nuclear and magnetic Bragg reflections, respectively.
		The two concentric ring-like features correspond to the (1 1 1) and (2 0 0) Debye-Scherrer rings from the copper sample holder.
		(b) and (c) Enlarged views of the reciprocal-space region enclosed by the  dotted square in Fig.~\ref{SENJU}(a), measured at 1.2 and 6 K, respectively.
		The superlattice reflection observed at 1.2 K appears at the same reciprocal position as at 0.3 K, whereas it is absent at 6 K.
		}
	\label{SENJU}
	\end{center}
\end{figure}
\begin{figure}[ht]
	\begin{center}
		\caption{
		One-dimensional cuts along (a) (1.7, 0, $l$) and (b) ($h$, 0, 0.66) through the magnetic Bragg peak near (1.7, 0, 0.66), measured in phase II (0.3 K).
		Solid line represents the results of Gaussian fit.
		}
	\label{QVec}
	\end{center}
\end{figure}
\begin{figure}[ht]
	\begin{center}
		\caption{
		(Color online) $\theta$-2$\theta$ scans through the magnetic Bragg peaks at (a) (2, 0, 2) - ${\bf q}$, (b) (2, 0, 2) + ${\bf q}$, and (c) (2, 0, -4) + ${\bf q}$ measured in phases I and II.
		The ratio of integrated intensities, $I_{\rm 1.2\ K}/I_{\rm 0.3\ K}$, is approximately 0.6 for all magnetic reflections. 
		}
	\label{IntegInt}
	\end{center}
\end{figure}
\begin{figure}[ht]
	\begin{center}
		\caption{
		(Color online) (a) ZF-$\mu$SR spectra of SmAu$_3$Al$_7$ measured at 5.1, 2.0, and 0.31 K.
		The solid curves represent the best fits using Eq.~(1).
		(b) Temperature dependence of the high-field cutoff of the continuous distribution of local magnetic fields at the muon site, $B_{\rm max}$.
		The normalized $B_{\rm max}$ values (right axis) are compared with the normalized square roots of the magnetic scattering intensities obtained from the neutron experiments.
		The solid line represents a fit based on the gap function expected for IC ordering.
		Both data sets scale well with the same curve.
		}
	\label{ZF_OP}
	\end{center}
\end{figure}
\setcounter{figure}{0}

\begin{figure}[ht]
	\begin{center}
		\includegraphics[width=\columnwidth]{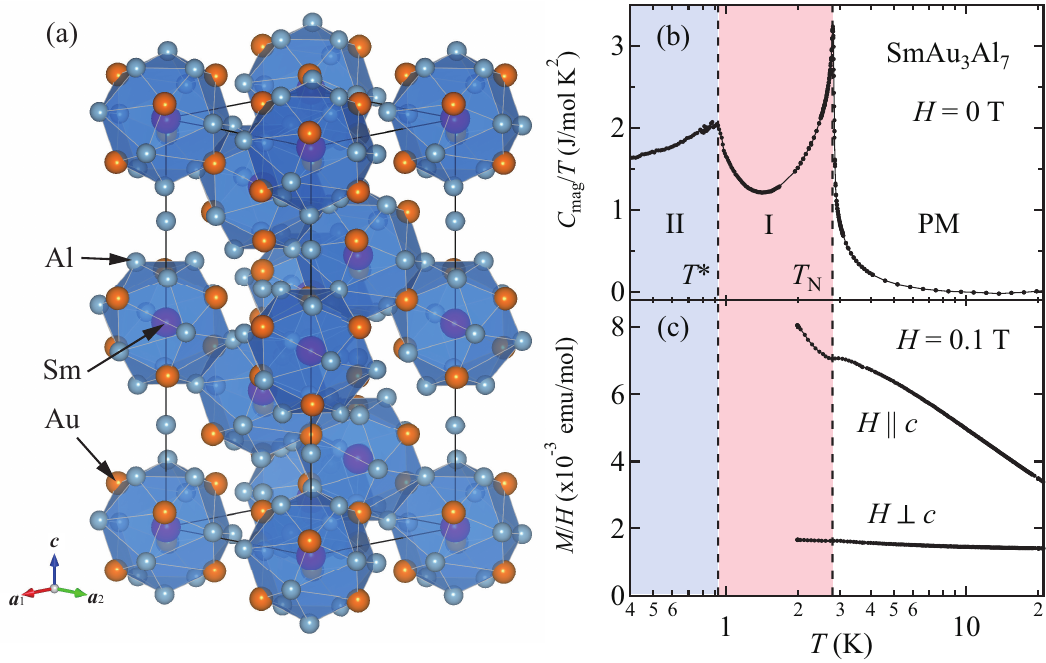}
		\caption{
		}
	\end{center}
\end{figure}
\begin{figure}[ht]
	\begin{center}
		\includegraphics[width=\columnwidth]{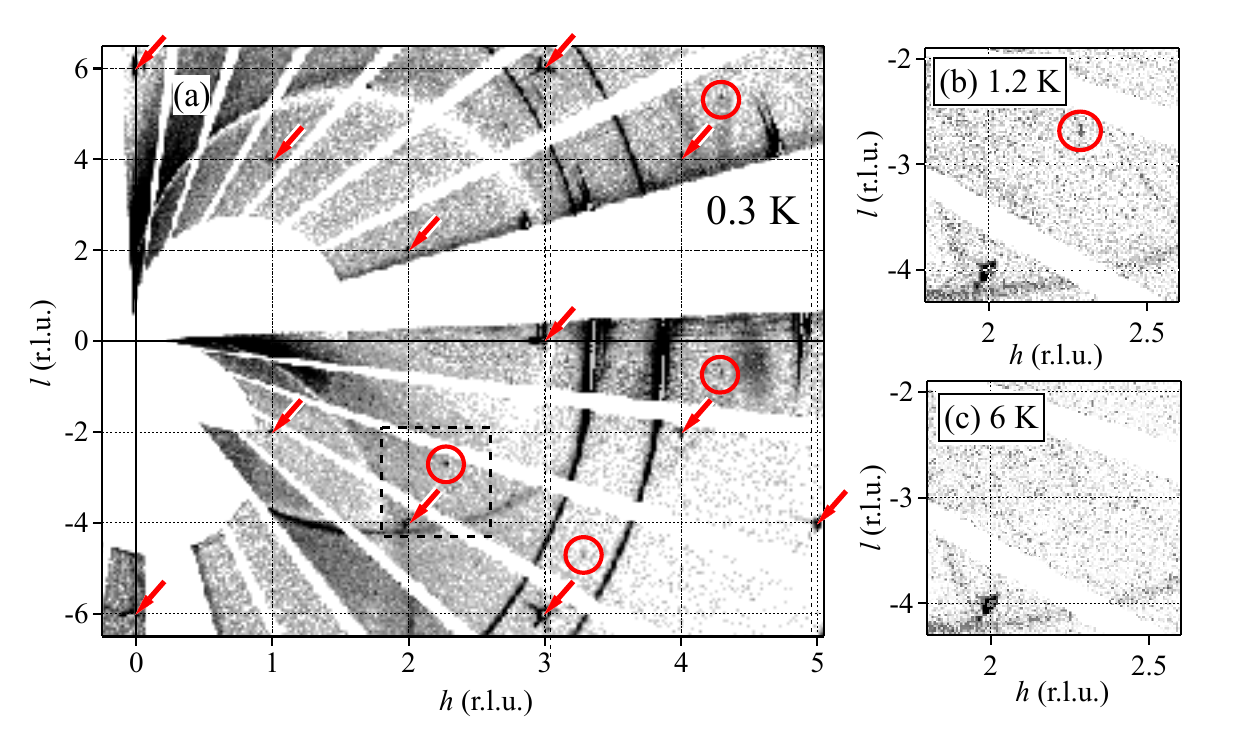}
		\caption{
		}
	\end{center}
\end{figure}
\begin{figure}[ht]
	\begin{center}
		\includegraphics[width=\columnwidth]{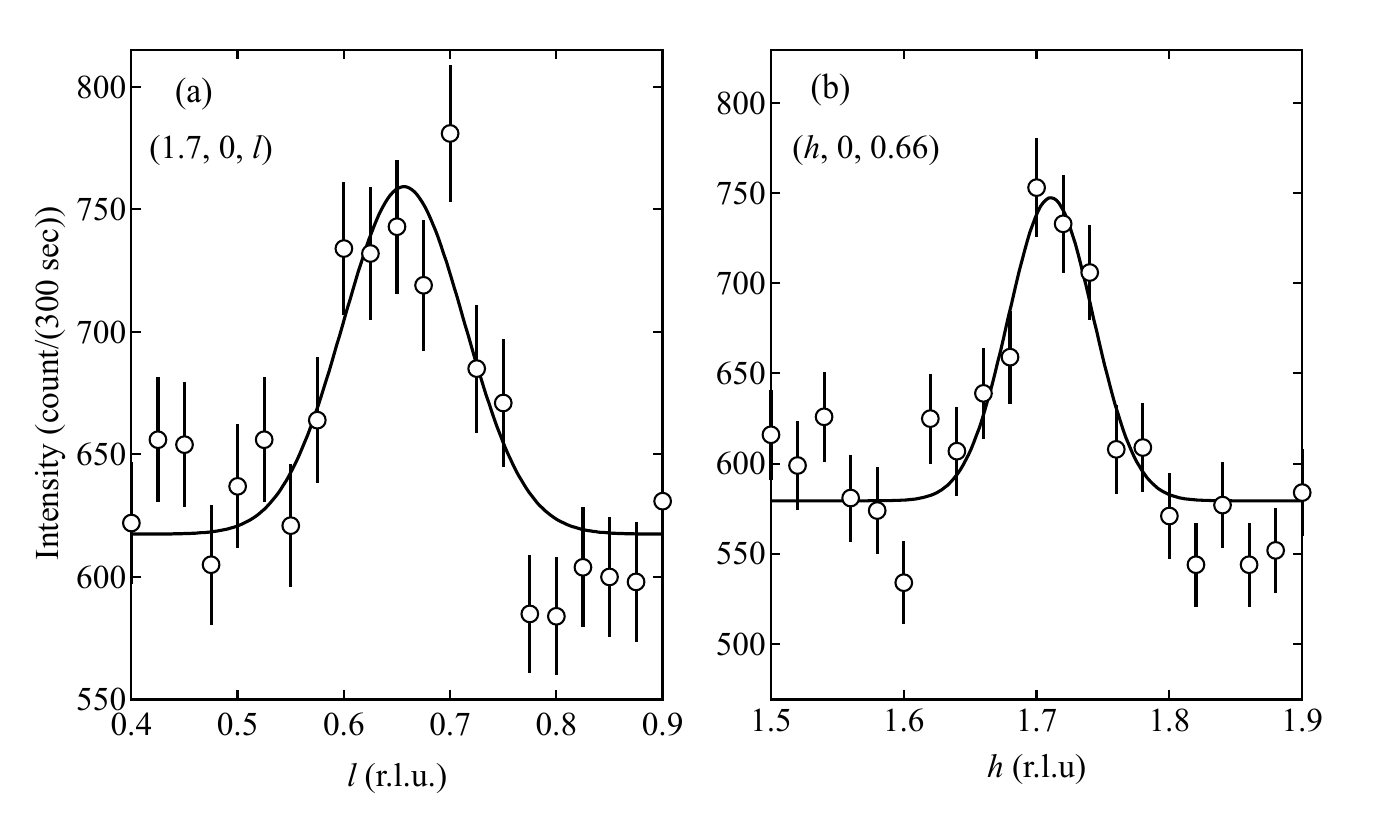}
		\caption{
		}
	\end{center}
\end{figure}
\begin{figure}[ht]
	\begin{center}
		\includegraphics[width=\columnwidth]{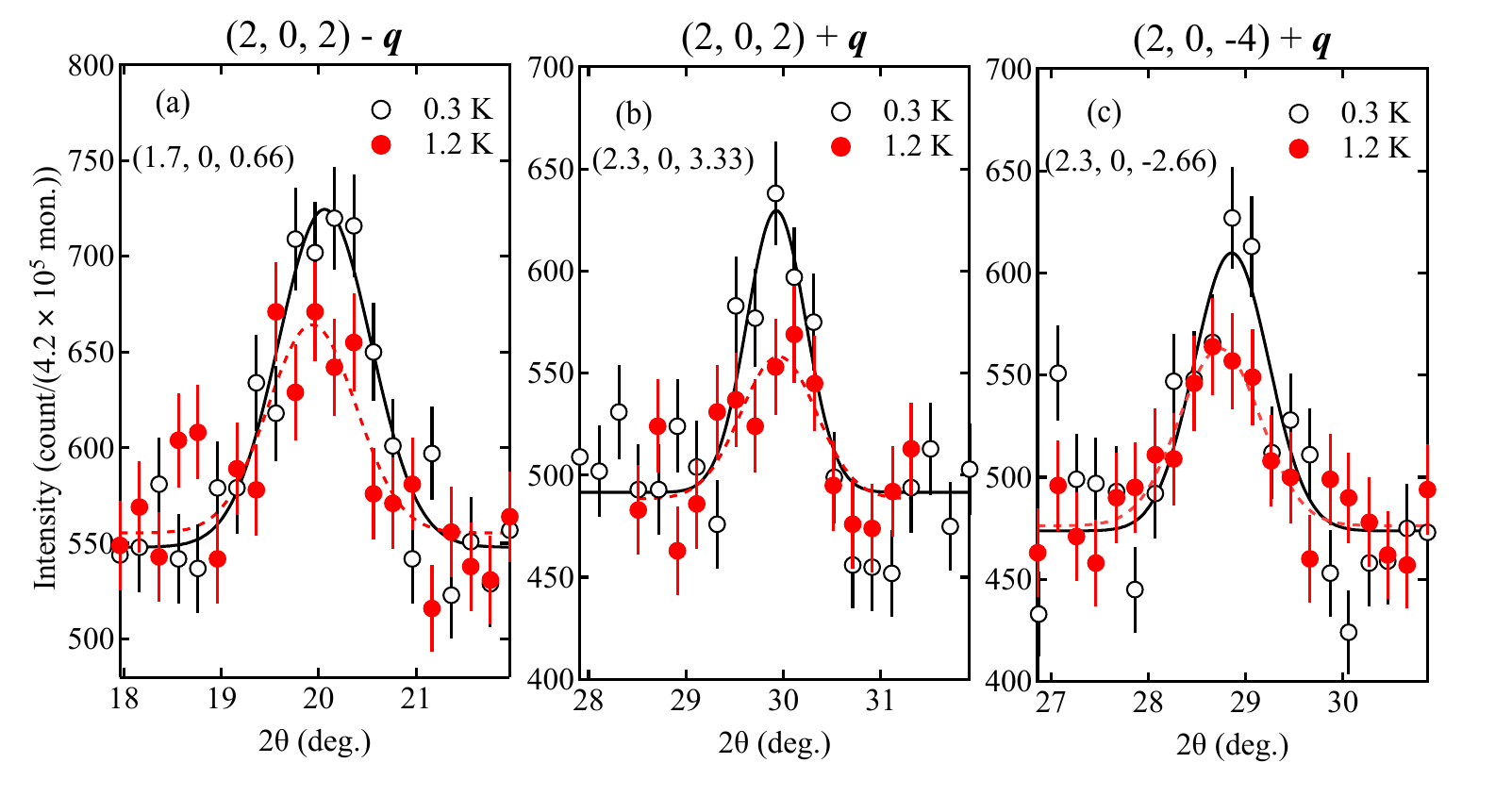}
		\caption{
		}
	\end{center}
\end{figure}
\begin{figure}[ht]
	\begin{center}
		\includegraphics[width=\columnwidth]{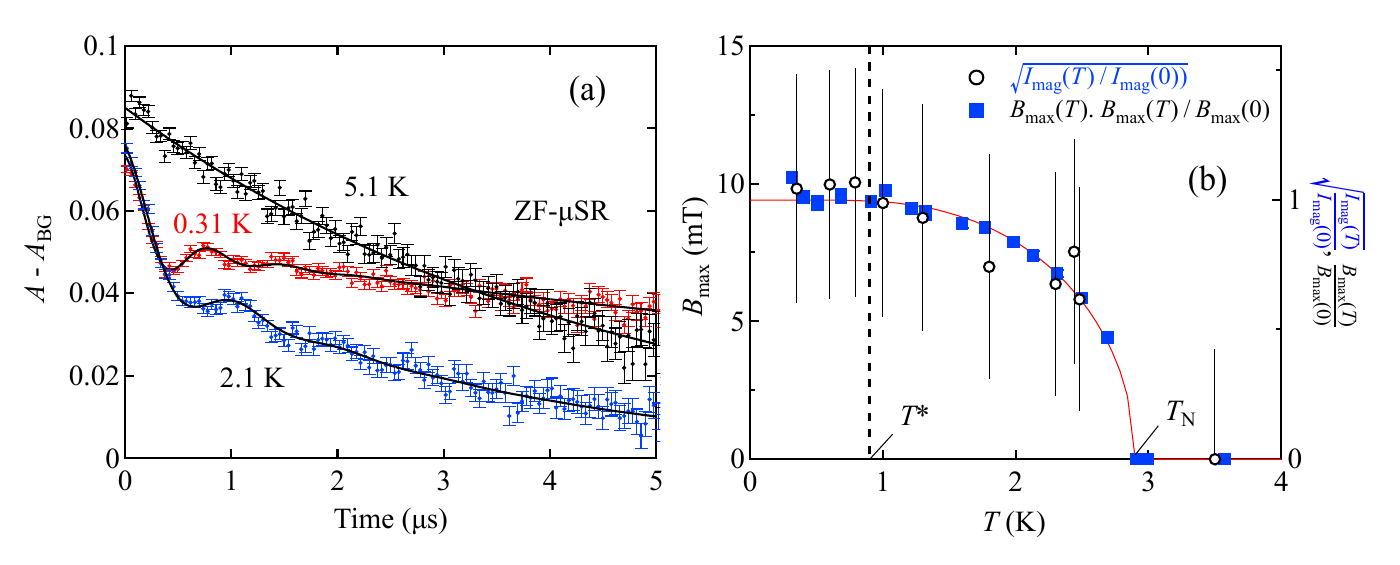}
		\caption{
		}
	\end{center}
\end{figure}
\end{document}